ORIGINAL PAPER

# Neuroglobin protects nerve cells from apoptosis by inhibiting the intrinsic pathway of cell death

Subhadip Raychaudhuri · Joanna Skommer ·
Kristen Henty · Nigel Birch · Thomas Brittain

**Abstract** In the past few years, overwhelming evidence has accrued that a high level of expression of the protein neuroglobin protects neurons in vitro, in animal models, and in humans, against cell death associated with hypoxic and amyloid insult. However, until now, the exact mechanism of neuroglobin's protective action has not been determined. Using cell biology and biochemical approaches we demonstrate that neuroglobin inhibits the intrinsic pathway of apoptosis in vitro and intervenes in activation of pro-caspase 9 by interaction with cytochrome $c$. Using systems level information of the apoptotic signalling reactions we have developed a quantitative model of neuroglobin inhibition of apoptosis, which simulates neuroglobin blocking of apoptosome formation at a single cell level. Furthermore, this model allows us to explore the effect of neuroglobin in conditions not easily accessible to experimental study. We found that the protection of neurons by neuroglobin is very concentration sensitive. The impact of neuroglobin may arise from both its binding to cytochrome $c$ and its subsequent redox reaction, although the binding alone is sufficient to block pro-caspase 9 activation. These data provides an explanation the action of neuroglobin in the protection of nerve cells from unwanted apoptosis.

## Introduction

Apoptosis, an evolutionarily conserved form of cell death, is one of the most complex signalling pathways known. There are two main initiation pathways of apoptosis: the death receptor (extrinsic) pathway and the mitochondrial (intrinsic) pathway. The latter can be triggered by cytotoxic stress that induces the activation of BH3-only and multi-domain pro-apoptotic Bcl-2 family members. In particular, certain BH3-only proteins, including tBid, directly activate multidomain Bcl-2 proteins Bax and Bak [1]. This event is negatively regulated by anti-apoptotic Bcl-2 proteins, such as Bcl-2, which can be inhibited by another set of BH3-only proteins referred to as indirect activators [1]. Activated Bax/Bak oligomerise and lead to mitochondrial outer membrane permeabilisation (MOMP), causing the release of pro-apoptotic intermembrane proteins including cytochrome $c$. Once in the cytosol, cytochrome $c$ in the presence of (d)ATP initiates the oligomerisation of Apaf-1 (apoptotic protease activating factor 1) into a large protein scaffold, the apoptosome, which recruits and processes pro-caspase 9 into an active initiator caspase 9. The initiator caspase 9 then activates the executioner caspases, such as caspase 3, which bring about the apoptotic destruction of the cell. This process is illustrated in Fig. S1.

Activation of effector caspases is considered as a point of no return in the apoptotic process. Therefore, the up-

S. Raychaudhuri (✉)
Department of Biomedical Engineering, University of California, Davis, 2521 Genome and Biomedical Sciences Bld., 451 Health Sciences Drive, Davis, CA 95616-5294, USA
e-mail: raychaudhuri@ucdavis.edu

J. Skommer (✉) · K. Henty · N. Birch · T. Brittain (✉)
School of Biological Sciences, University of Auckland, Thomas Bld., 3a Symonds Street, Auckland 1142, New Zealand
e-mail: J.Skommer@auckland.ac.nz

T. Brittain
e-mail: T.Brittain@auckland.ac.nz

stream events such as MOMP and apoptosome formation need to be tightly regulated to ensure that the threshold for cell death has been reached and unwanted apoptosis will not occur [2]. Tight control of apoptosis is particularly advantageous for postmitotic cells such as neurons which have limited regenerative potential and need to last for the life of the organism. It has been previously suggested that the strict regulation of apoptosis in neurons is mediated by the inhibitor of apoptosis family of proteins (IAPs) and/or low expression of Apaf-1 [3]. The protein neuroglobin, first discovered by Burmester [4] during a bioinformatics search of the human genome for previously undiscovered heme proteins, also protects neural cells from a variety of apoptotic insults. The heme protein neuroglobin is found at high levels in certain neurons, endocrine cells and retinal rod cells, where it can reach a local concentration of 100 μM and is closely associated with mitochondria [5]. Over recent years neuroglobin has been shown to provide a protective function with respect to amyloid toxicity [6] and Alzheimer's disease [7], whilst also protecting from cell death associated with hypoxic ischemia both in vitro and in vivo [8–11]. All these processes are at least in part associated with apoptotic cell death. This indicates that neuroglobin may provide until now unidentified means of apoptosis regulation in neurons [12].

Neuroglobin is an iron containing heme protein, and as such it can exist in either a ferrous or ferric form and can also take part in redox reactions with other redox active proteins, such as cytochrome $c$. The neuroglobin-cytochrome $c$ complex can undergo a redox reaction, where ferric cytochrome $c$ is reduced to a non-apoptotic ferrous form [13], lending argument for the hypothesis that the protective role provided by neuroglobin could arise from its intervention in the mitochondrial pathway of apoptosis.

In the current study we use a combination of experimental studies and a computational systems biology approach to investigate the role of neuroglobin in regulation of the mitochondrial pathway of apoptosis. Neuroglobin prevents cytochrome c-induced caspase 9 activation, significantly blocking cell death triggered by a BH3 mimetic HA14-1. Computational analysis provides a quantitative model of neuroglobin inhibition of apoptosis at a single cell level, which is supported by our experimental data.

## Materials and methods

### Cell culture and reagents

Human neuroblastoma SH-SY5Y cells were grown in advanced DMEM/F12 (1/1) medium (Invitrogen) supplemented with L-glutamine, 1% penicillin/streptomycin and 10% fetal bovine serum (Invitrogen) at 37°C in humidified 95% air, 5% $CO_2$. Jurkat T cells were grown in RPMI 1640 medium (Invitrogen) supplemented as above. To induce cell death, SH-SY5Y cells were treated with a small-molecule BH3 mimetic HA14-1 (Alexis Biochemicals).

### Plasmids, transfections, stable cell lines

Vector expressing human neuroglobin was generated using a human cDNA clone from Origene (Rockville, MD, USA) subcloned into the KpnI and XbaI sites of pcDNA3.1 and the construct verified by direct sequencing. Plasmids were propagated using DH5α competent cells, according to standard protocols, and purified using EndoFree Maxi Prep kits (Qiagen). SH-SY5Y cells ($1 \times 10^6$) were plated on 10 cm tissue culture dishes and transfected with pcDNA3.1 or NGB_pcDNA3.1 using calcium phosphate transfection, according to standard protocols. 16 h post-transfection cells were washed 3 times with serum-free medium, and cultured in fresh medium for 48 h to allow for transgene expression. The medium was then replaced with selection medium containing G418 (400 μg/ml; Invitrogen). Eight to twelve clones from each transfection were selected following 4 weeks of selection. Based on screening for neuroglobin expression using a mouse monoclonal anti-neuroglobin Ab (13C8) (Abcam, Cambridge, UK) three stable clonal cell lines were selected for further analysis.

### qRT-PCR

Total RNA was isolated from both adherent and floating cells using TRIzol (Invitrogen), according to manufacturer's instructions. cDNA synthesis was performed using a High Capacity cDNA Reverse Transcription Kit with RNase Inhibitor (Applied Biosystems), according to manufacturer's instructions. Quantitative RT-PCRs were performed using 10 ng (RNA equivalents) of cDNA as a template. Gene-specific primers and probes, and TaqMan Gene Expression Master Mix were from Applied Biosystems and amplifications were performed using an ABI Prism 7900 instrument. Data were normalized to 18S rRNA expression.

### Western blotting

Total cell lysates were prepared using RIPA buffer supplemented with protease inhibitor cocktail (Roche). Proteins were separated using 15% gels and transferred onto PVDF membrane. The membranes were blocked with 5% non-fat milk in Tris-Buffered Saline-Tween 20 (TBS-T) for 45 min, followed by incubation with mouse monoclonal anti-neuroglobin antibody 13C8 (Abcam, Cambridge, UK), goat polyclonal anti-neuroglobin antibody E-16 (SantaCruz), or mouse monoclonal anti-tubulin antibody at 1:1,000, 1:100 and 1:9,000 dilution, respectively, using 1% milk in TBS-T. Low-salt TBS-T was used with monoclonal antibodies. Next, the membranes were washed 4 times in TBS-T, incubated for 45 min with anti-mouse (BioRad) or anti-goat (Jackson ImmunoResearch Laboratories, West Grove, PA, USA) HRP-conjugated secondary Abs, washed again 4 times in TBS-T, and developed using ECL reagent (GE Healthcare).

Flow cytometry

To assess cell viability, $1.5 \times 10^5$/well cells were plated on 24-well plate, and treated on the next day as indicated (legend Fig. 1). At the end of the experiment, both floating and adherent cells were collected, washed with PBS, stained for 5 min with propidium iodide (PI; 1 μg/ml) (Invitrogen), and analyzed immediately on FACS Calibur (BD). Cells were also gated based on forward scatter (FSC) and side scatter (SSC) to exclude cell debris, and next analyzed based on PI fluorescence using CellQuest (BD).

Colony forming assay

pcDNA3.1 and NGB_pcDNA3.1 stable SH-SY5Y cell lines were plated at $5 \times 10^4$ cells/well in 12-well plates. The next day the cells were treated with DMSO or HA14-1 (12.5 μM) for 48 h. Cells were then washed, and cultured in fresh medium. Cells were cultured over 10–12 days in DMEM/F12 supplemented with 10% FBS, penicillin/streptomycin, glutamine and G418 (400 μg/ml; Invitrogen). Medium was replaced every 3 days. Cell colonies were stained with Giemsa, fixed in 10% methanol, photographed, and then counted using ImageJ software.

Bacterial expression and protein purification

Bacterial expression of neuroglobin followed our previously published protocols [14]. The gene coding for human neuroglobin was cloned into pET29a. *E. coli* BL21 (DE3) was then transformed with the pET plasmid and grown for 2 days at 28°C under micro-aerobic conditions without induction in TB medium supplemented with 10% phosphate buffer and kanamycin. Cells were harvested by centrifugation at 5,000$g$ for 10 min. The cells were next washed and then ultrasonically lysed in 50 mM Tris buffer containing 1 mM EDTA and 5 mM DTT at pH 8.0 (lysis buffer). The cell lysate was centrifuged at 20,000$g$ for 10 min and the supernatant sequentially raised to 30 and then 60% ammonium sulphate. The final precipitate, obtained by centrifugation at 20,000$g$ for 20 min, after being redissolved in 50 mM Tris buffer containing 1 mM EDTA and 5 mM DTT was then extensively dialysed

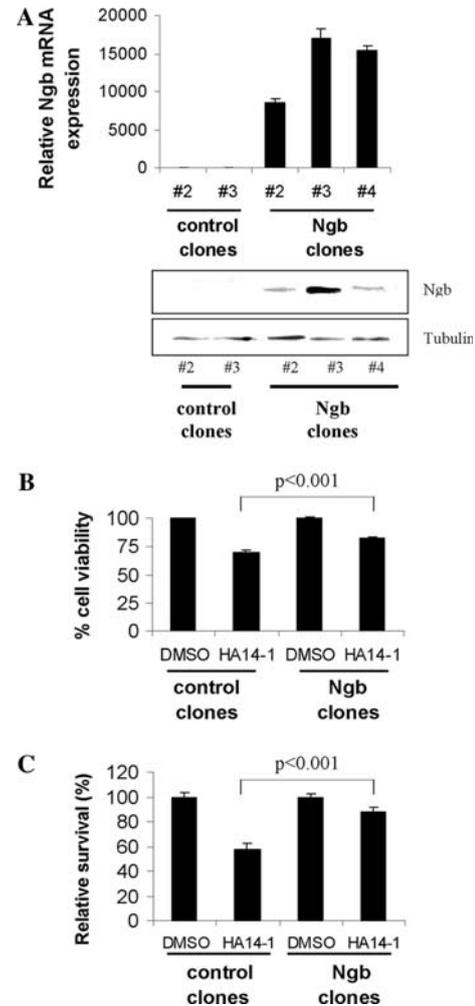

**Fig. 1** Over-expression of neuroglobin reduces sensitivity towards BH3 mimetic-induced cell death. **a** qRT-PCR (*upper panel*, $n = 3$) and western blot (*lower panel*) analysis of neuroglobin (Ngb) in stable SH-SY5Y-derived cell lines. **b** Flow cytometry analysis reveals that neuroglobin over-expression partially protects against cell death induced by 2.5 h exposure to 25 μM HA14-1 (mean ± SD, $n = 6$). Results represent the average from 2 control (pcDNA3.1) cell lines (#2 and #3) and 3 Ngb-over-expressing cell lines (#2, #3, and #4). **c** Long-term colony survival of wild-type and neuroglobin-over-expressing cells following 48 h treatment with 12.5 μM HA14-1 reveals nearly complete protection by neuroglobin (mean ± SD, $n = 10$)

against the same buffer. The dialysed solution was chromatographed for 120 min on Q Sepharose using XK26 column (Amersham Biosciences) and a linear gradient of 100% lysis buffer and 100% lysis buffer with 200 mM NaCl. Next, the solution was gel filtered using Superdex 75 column (Amersham Biosciences; 10/300 GL) using lysis buffer with 150 mM NaCl to elute the sample. Flow rate on both columns was 1 ml/min. Taking advantage of the hyper-thermal stability of neuroglobin [15], between columns the neuroglobin containing fraction was heated to 75°C for 3 min to remove contaminating proteins. We have determined the x-ray structure of protein produced by this method and shown that it is identical to that published for neuroglobin [16]. The final product purity was assessed by PAGE and electrospray mass spectral analysis. The form of neuroglobin employed had Cys-Ser mutations and so avoided any confounding effects of S–S formation.

### Pro-caspase 9 activation

Activation of pro-caspase 9 in the presence of various forms of neuroglobin was assessed using a caspase 9 assay recently established in our laboratory. This assay uses the cytosol obtained from cultured Jurkat cells to provide the major components of the apoptosome (dATP, Apaf-1 and pro-caspase 9) [17–19]. Cells, grown as outlined above, were harvested by centrifugation, washed twice with ice-cold PBS, and immediately resuspended in four volumes of ice cold lysis buffer (250 nM sucrose, 10 mM KCl, 20 mM HEPES, pH 7.4, 1.5 mM $MgCl_2$, 0.5 mM EDTA, 0.5 mM EGTA, and protease inhibitor mixture (Roche). The resuspended cells were immediately dounced with a homogenizer (Thomas PHILA, USA) and Teflon pestle and the resulting homogenate was centrifuged at 13,200 rpm for 30 min at 1°C. Activation of pro-caspase 9 was initiated by the addition of ferrous or ferric cytochrome $c$ (6 μM), either in the absence or presence of added ferric neuroglobin (6 μM), to the cell lysate. Following an incubation period of 30 min the activation of pro-caspase 9 was monitored by following the cleavage of an added luminescent peptide substrate for caspase 9 (Glo 9, Promega). The emitted light was measured using a Wallac EnVision 2104 multilabel reader (Perkin Elmer).

### Statistical analysis

Differences between the groups were determined by ANOVA and $t$-test using SPSS Statistics 17.0.

### Computational systems analysis

We use a kinetic Monte Carlo technique to simulate the signalling reactions in the intrinsic apoptotic pathway. Details of our simulation method can be found in our earlier work [20, 21]. Below, we provide a brief outline of our simulation scheme describing (1) the Monte Carlo sampling of signaling reactions in the intrinsic apoptotic pathway, and, (2) how the effect of neuroglobin on the intrinsic apoptotic pathway is incorporated in our model. In our Monte Carlo (MC) algorithm, diffusion and reaction are treated in an equivalent manner by defining probabilistic rate constants for both events. Individual molecules are sampled randomly and two possible moves are considered—(a) diffusion move, and, (b) reaction move. The diffusion probability of multi-molecular complexes is taken to be lower than that of free molecules. For example, the apoptosome complex, made out of two molecules of cytochrome c and two molecules of Apaf-1, is considered immobile. Explicit simulation of individual molecules allows us to incorporate mutual exclusion of molecules from occupying the same spatial location. When a molecule is chosen for reaction it can undergo any of the several reactions possible for that molecular species. Some general form of reactions are: (1) two free molecules form a complex with probability $P_{on}$, (2) a molecular complex can dissociate with probability $P_{off}$, (3) catalysis reaction such as proteolitic cleavage of a molecule during dissociation from a molecular complex ($P_{cat}$). Formation of multimolecular complexes such as the apoptosome is handled by assigning binding pockets with a molecule like cytochrome c and then defining explicit rules of binding and dissociation at different binding pockets.

### Neuroglobin-cytochrome $c$ reaction

Neuroglobin can bind to cytochrome $c$ (Fig. S1) and prevent apoptosome formation. We assume the following rate constants for this neuroglobin binding to cytochrome $c$ reactions: $k_{on} = 10^7$ $M^{-1}$ $s^{-1}$ with an overall affinity constant $K_A = 2 \times 10^4$ $M^{-1}$ [22].

This simulation is carried out using a three-dimensional cubic (lattice) box that mimics a cell with signaling molecules inside it. In our simulations individual molecules are placed on individual nodes of the simulation box. At the beginning of the simulation we randomly place all the molecules either on the cell surface or inside the three-dimensional simulation box. We have used a previously developed mapping scheme between probabilistic parameters of our model and their physical analogs [20, 21]. Most of the parameter values, such as kinetic rate constants and molecular concentrations, used in our simulations are obtained from [23–27].

## Results

### Over-expression of neuroglobin protects SH-SY5Y cells against HA14-1-induced cell death

We have previously hypothesised that the protective role provided by the presence of neuroglobin could arise from its intervention in the mitochondrial pathway of apoptosis [12]. In order to study the anti-apoptotic effects of high neuroglobin levels we established stable neuroglobin (Ngb)-overexpressing human neuroblastoma SH-SY5Y cells (Fig. S2). The overexpression of Ngb was confirmed using quantitative RT-PCR and western blotting (Fig. 1a).

The pcDNA3.1 and Ngb-overexpressing cell lines were treated with DMSO or HA14-1, a BH3 mimetic which inhibits anti-apoptotic Bcl-2 protein(s) and thus specifically activates the intrinsic pathway of apoptosis. Using FACS (fluorescence activated cell sorting) analysis we show that Ngb-overexpressing cells have an enhanced resistance to cell death induced by HA14-1. Neuroglobin over-expressing cells exposed to HA14-1 (25 μM for 2.5 h) show a significantly lower rate of cell death than do wild-type cells during the first hours of treatment (Fig. 1b). Furthermore, the protective action of neuroglobin was reflected in terms of long-term survival, as it was found to significantly increase the clonogenic potential of SH-SY5Y cells following prolonged (12.5 μM for 48 h) exposure to HA14-1 (Fig. 1c). These observations strongly support our hypothesis that neuroglobin exerts its protective role by intervening in the mitochondrial pathway of apoptosis.

## Computational analysis of the cytochrome $c$-neuroglobin complex

We have previously observed that neuroglobin reacts very rapidly with cytochrome $c$ [12, 13, 22]. Computational analysis using *BiGGER* [28] has identified a lowest energy, putative structure for the cytochrome $c$-neuroglobin complex formed in this reaction (Fig. 2). The analysis of the interface amino acids of this structure suggests the binding of cytochrome $c$ Lys residues 72 and 25 to Glu residues 60 and 87 on neuroglobin. These interactions are likely to be of functional significance as the Lys residues 72 and 25 of cytochrome $c$ have previously been identified as mandatory and important for the binding of cytochrome $c$ to Apaf-1, respectively [29–32]. Furthermore, their binding partners on neuroglobin, namely Glu60 and Glu87, are conserved in the amino acid sequence from frog to human.

## Systems level modelling of the intrinsic pathway of apoptosis indicates that formation of the cytochrome $c$-neuroglobin complex suppresses apoptosis

The complex nature of protein signaling networks, such as apoptotic pathways, with multiple variables acting at the same time, can be analyzed using a systems level probabilistic approach. Therefore, using systems level information of the apoptotic signalling reactions, we have developed a quantitative model of neuroglobin-mediated inhibition of apoptosis, which probes the death-protective role of neuroglobin at a single cell level [21]. This method is based on a probabilistic Markov chain method in which the reactivity of all the signalling molecules follows a stochastic, rather than deterministic behaviour [21]. In our computational model, the mitochondrial pathway of apoptosis is initiated by caspase 8-induced activation of a BH3-only protein, Bid, to tBid. tBid further interacts with a set of pro- and anti-apoptotic proteins leading to activation of Bax and its translocation to the mitochondrial surface (Fig. S1). The activation of Bax in our model is represented by tBid-Bax and Bax2 complexes, which simulate Bax oligomerisation induced directly by BH3-only proteins and by general inhibition of anti-apoptotic Bcl-2 proteins (Fig. S1). In our computational model, tBid and Bax simulate the effect of all the BH3-only and multi-domain pro-apoptotic Bcl-2 proteins, respectively, whereas all the anti-apoptotic effects are represented by Bcl-2. Mitochondrial cytochrome $c$ is released into the cytosol once Bax activation reaches a critical threshold determined by a preset number of Bax2 complexes [33, 34]. The intrinsic stochastic nature of the signalling reactions up-stream of mitochondria in the intrinsic pathway of apoptosis makes the Bax2 activation a stochastic event. The loop network structure of tBid-Bcl-2-Bax (Fig. S1) and the Bax/Bcl-2 ratio are the critical determinants of MOMP and cytochrome $c$ release. In our model cytochrome $c$ is released in an all-or-none manner. This is in line with recent experimental studies suggesting that the release of cytochrome $c$ occurs in a rapid (~5 min) manner once it is initiated, and that this behaviour is cell type independent [35, 36]. The rapid release of cytochrome $c$ is initiated by stochastic Bax2 activation and therefore it is also a stochastic event, with cell-to-cell variation arising directly from its stochastic nature [21, 27]. Our model assumes also that cytochrome $c$, once released into the cytosol, binds with Apaf-1 and dATP to form the apoptosome (Fig. S1), with subsequent activation of caspase 9 and caspase 3 [21]. Finally, we use a lower amount of Bcl-2 in our simulations to mimic the effect of Bcl-2 inhibitors such as the BH3-mimetic HA14-1 that was used in our initial experiments (Fig. S1). Each run of our Monte Carlo simulation represents activation of apoptosis signalling through the intrinsic pathway at a single cell level. Interestingly, the time scale of the apoptosis through the intrinsic pathway emerges naturally from our stochastic simulations that show slow (∼hours) activation, as also observed in our experimental studies.

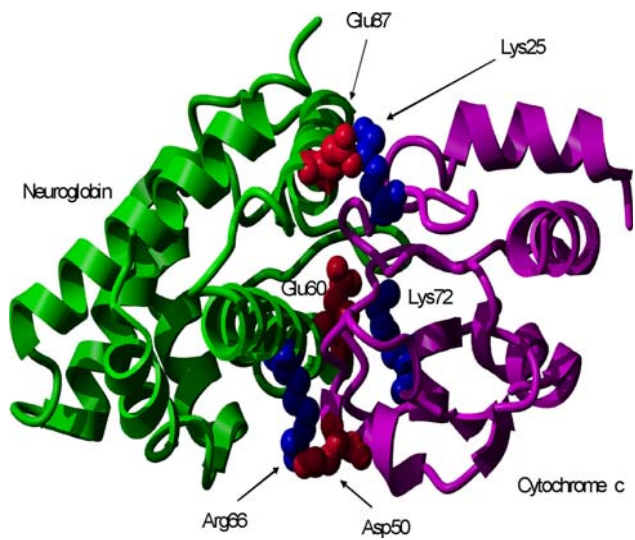

**Fig. 2** Lowest energy soft-docked structure of the complex between neuroglobin (*left*) and cytochrome $c$ (*right*) calculated using *BiGGER* [24] and rendered using YASARA (www.yasara.org)

In our computational model ferrous neuroglobin binds with ferric cytochrome $c$ and further changes it to a non-apoptotic reduced form [13], thus adding to its inhibitory effect on apoptotic signalling. When a small of amount of neuroglobin (nM concentrations) is introduced in our simulations it partially inhibits apoptosome formation by blocking cytochrome $c$-Apaf-1 binding. Hence, the activation of caspase-9 and caspase-3 are also slowed down with larger cell-to-cell stochastic fluctuations. Increasing the concentration of neuroglobin exerts a strong effect on apoptotsome formation and thus apoptosis inhibition. We analysed the time-course for (1) the first apoptosome formation, (2) caspase 9 activation, and (3) caspase 3 activation, as we varied the concentration of neuroglobin (Fig. 3). Clearly, the time-scale of apoptosome formation (Fig. 3, crosses) correlates well with the subsequent activation of caspase 9 (Fig. 3, lines) and caspase 3 (Fig. S2) at a single cell level. The time course of apoptosome formation is stochastic, with large cell-to-cell variations, that arise due to three major reasons: (1) release of cytochrome $c$ into the cytosol is a rapid stochastic event; (2) the intrinsic kinetic rate constant of cytochrome $c$-Apaf-1 complex formation is small; and (3) the bringing together of multiple pairs of cytochrome $c$ and Apaf-1 molecules to the same spatial location, by diffusive transport, is a low probability event. Once a few molecules of apoptosome are formed, activation of caspase-9 and caspase-3 occurs rapidly. Thus, the fluctuation in the time-course of apoptosome formation is translated into large fluctuations in caspase-9 and caspase-3 activation, associated with a sharp spike-like rise of activated caspases, at a single cell level. Figure 3 shows how increasing neuroglobin levels lead to slower and lesser apoptosome formation, in a given simulation time scale; due to inhibition of the low probability event of apoptosome formation. Neuroglobin concentrations in excess of 1 μM have been reported in mouse brain and retinal cells [5, 37]. We also calculate the probability distribution of casapse 3 activation in order to characterize cell-to-cell stochastic fluctuations in apoptosis signalling under neuroglobin inhibition. The probability of caspase 3 activation shows a characteristic bi-modal distribution, over time, that changes due to neuroglobin addition to the system (Fig. 4; [21]).

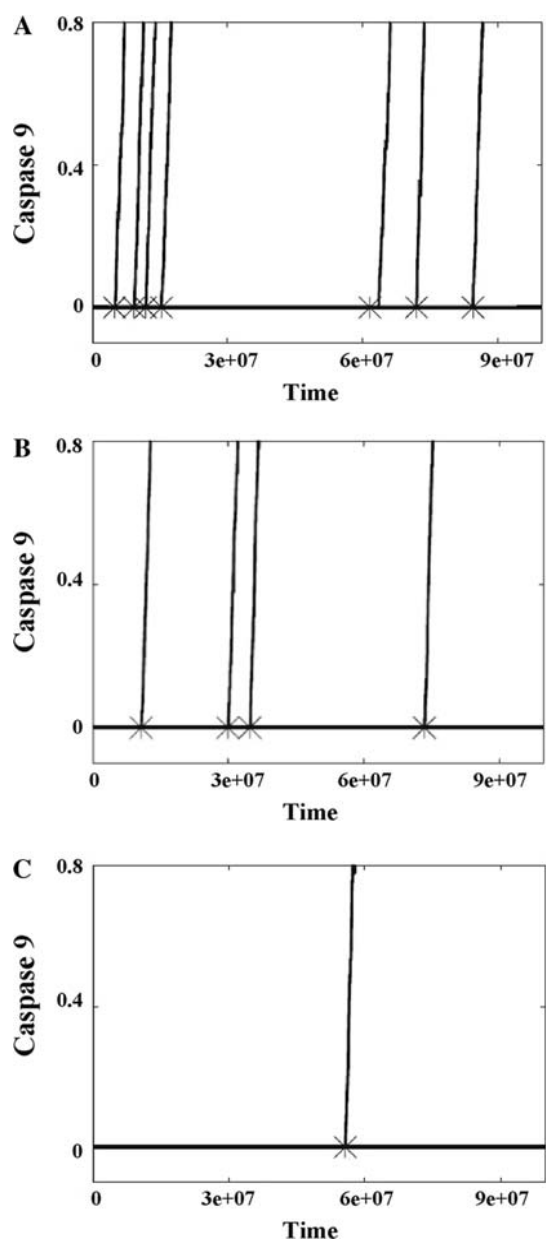

**Fig. 3** The activation of apoptosome and caspase-9 depend on neuroglobin concentration. **a** Neuroglobin = 0, **b** neuroglobin = 0.01 μM, and **c** neuroglobin = 0.1 μM. Cytochrome $c$ concentration is uniform at 0.1 μM in all three studies. Time is measured in monte carlo (MC) simulations steps. 1 MC step = $10^{-4}$ s, hence time-scale shown $10^8$ MC steps ∼2.8 h. Caspase 9 activation is shown (up to 80% completion) normalized by the maximum. Each cross and line represent apoptosome formation and caspase 9 activation, respectively, at a single cell level

Redox reaction between cytochrome $c$ and neuroglobin is not necessary for the anti-apoptotic action of neuroglobin, but may modulate it

Previous studies [22] have indicated that ferric neuroglobin binds to ferric cytochrome $c$ with an equilibrium constant in the micro-molar range. Our experimental data indicate

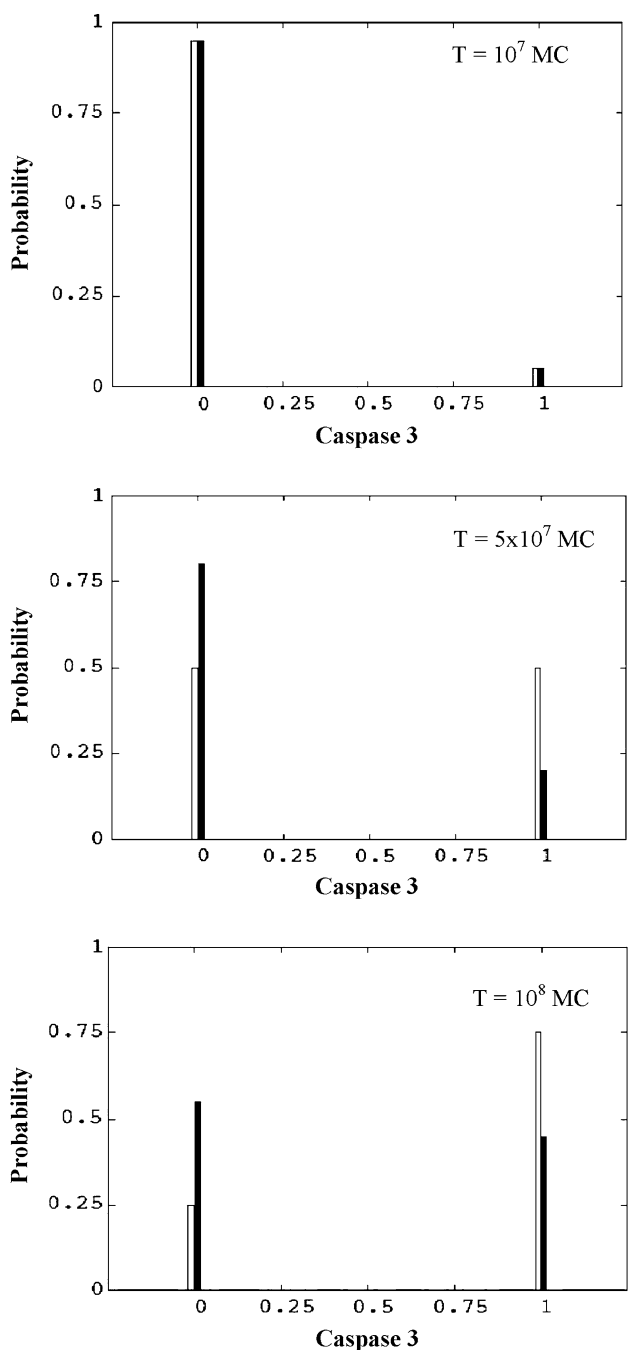

**Fig. 4** Probability distribution of caspase 3 calculated from single cell activation data of caspase 3. We show results for two neuroglobin concentrations: neuroglobin = 0 (*white bars*) and neuroglobin = 0.01 µM (*black bars*). Cytochrome $c$ concentration is fixed at 0.1 µM. Caspase 3 concentration (*x*-axis) is normalized by the maximum. Time is measured in monte carlo (MC) simulations steps. 1 MC step = $10^{-4}$ s

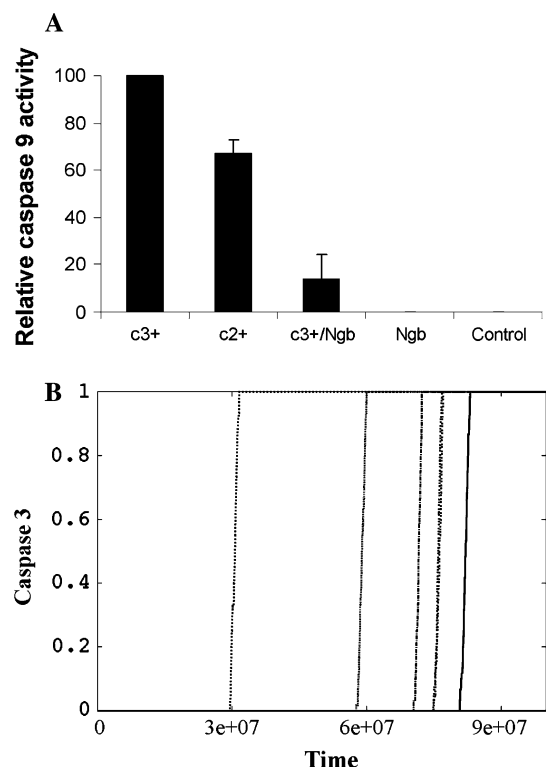

**Fig. 5** Neuroglobin binding to cytochrome $c$ is sufficient to block activation of caspase 9. **a** Assay of the apoptosome activation of caspase 9 used the cytosol obtained from cultured Jurkat cells to provide the major components of the apoptosome (dATP, Apaf-1 and pro-caspase 9). Apoptosome activation was initiated by the addition of cytochrome $c$ either in the absence or presence of added neuroglobin, and the subsequent assembly of the apoptosome monitored by following the cleavage of a luminescent peptide substrate for caspase 9. **b** Caspase 3 activation data without the neuroglobin reduction of cytochrome $c$ to a non-apoptotic form. We used cytochrome $c$ 0.1 µM and neuroglobin 0.1 µM in these simulations. Each line shows the time-course of caspase-3 activation at the single cell level. Caspase 3 activation is normalized by the maximum. Time is measured in monte carlo (MC) simulations steps. 1 MC step = $10^{-4}$ s

that ferric neuroglobin drastically reduces the apoptotic activity of ferric cytochrome $c$, whilst itself has no impact on activation of pro-caspase 9 (Fig. 5a). We also find, in this reconstituted in vitro system, that caspase 9 activation by exogenous cytochrome $c$ depends on the redox state of the added cytochrome $c$ (Fig. 5a). This suggests that the redox-reaction between cytochrome $c$ and neuroglobin, even if it is not an absolute requirement, modulates the neuroglobin-mediated inhibition of cytochrome $c$-induced caspase 9 activation. In order to determine the relative contributions of the binding and redox reactions of neuroglobin we have also simulated the alternative mechanism of apoptosis inhibition by neuroglobin, under the assumption that ferric neuroglobin acts by binding with ferric cytochrome $c$ but does not reduce the bound cytochrome $c$. In such a scenario, the high dissociation constant of the cytochrome $c$-neuroglobin complex allows more frequent association of cytochrome $c$ with Apaf-1, resulting in increased chances of apoptosome formation and thus increased apoptosis (compare Figs. 3c, 5b). Hence, a higher level of cytosolic neuroglobin is needed to completely abolish apoptotic activation in the absence of redox reaction. However, the qualitative changes due to variation in neuroglobin concentration, such as slower apoptotic activation with large cell-to-cell variation, as seen in Fig. 3, remain unaltered (Fig. 5b).

### Increased concentration of cytosolic cytochrome *c* requires higher levels of neuroglobin to block apoptosis

We have also modelled the effect of a higher level of cytochrome *c* release, upon Bax reaching the preset threshold activation level. The mitochondria typically contain sufficient cytochrome *c* to produce a final concentration of approximately 0.5 μM within the cell, if completely released. Experimental studies, carried out under controlled conditions, however have shown that 20% of cytochrome *c* ($\sim$0.1 μM) within mitochondria is free in the inter-membrane space of the mitochondrion, which can be released upon mild apoptotic challenge [38] whilst stronger apoptotic stimuli can result in a higher dose of cytochrome *c* release from mitochondria. We used our computational model to systematically study the effect of increasing cytochrome *c* concentration (released into cytosol) on apoptotic activation. The activation of caspase 3 in our model system is found to be very sensitive to the ratio of neuroglobin to cytochrome *c*; increasing >15 fold when cytosolic cytochrome *c* concentration is increased from 0.1 to 0.2 μM, at a fixed 0.1 μM concentration of neuroglobin (Fig. 6). In these studies we included a cytochrome c-neuroglobin complex dissociation step in our simulations, in addition to the redox conversion reaction, in order to better access the amount of neuroglobin needed to block apoptosis. At a cytochrome c concentration of $\sim$0.2 μM, an equal concentration of neuroglobin was unable to protect cells from apoptosis, but 0.5 μM neuroglobin was protective. Thus an increasingly higher level of neuroglobin is needed to block apoptotic signalling, which is affected by increases in cytochrome *c* concentrations in a non-linear manner. Our modelling data is consistant with the level of neuroglobin found in nerve cells having the capacity to protect the cells from mild challenge, such as those that might arise from calcium fluctuations encountered during normal nerve function, but permits the initiation of apoptosis on receipt of a stronger apoptotic signal. Again, this is in line with our experimental data which shows that the level of neuroglobin expression reached in our cell culture system (approx. 5 μM) confers better protection against low-level mitochondrial stress than high-level stress (Fig. 1b, c and Fig. S3).

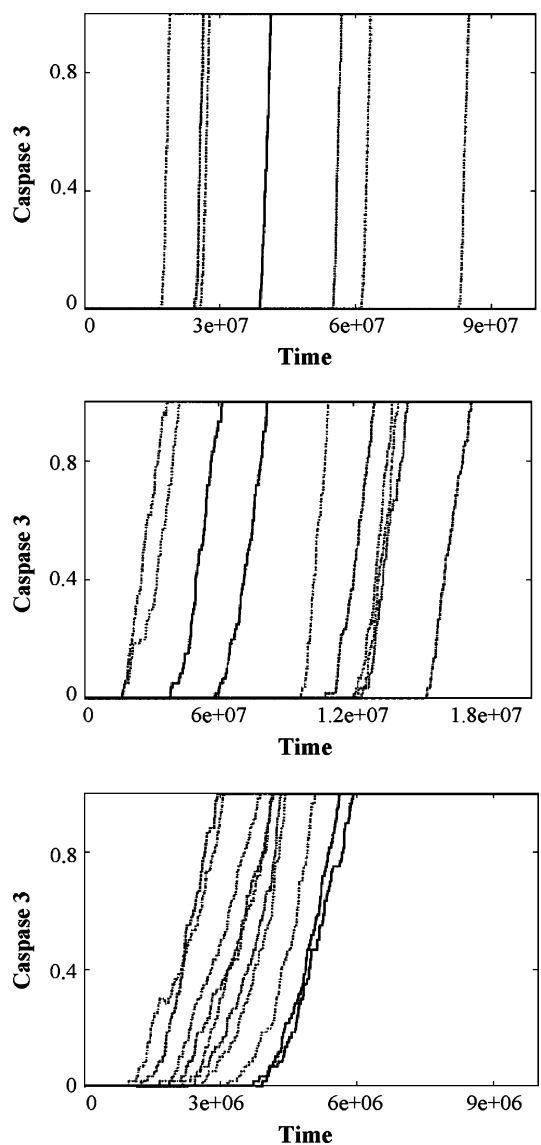

**Fig. 6** Increased levels of cytochrome *c* require higher levels of neuroglobin to block activation of caspase 3. Time course of caspase-3 activation is shown for three different values of cytochrome *c* concentrations: **a** 0.2 μM, **b** 0.5 μM, **c** 1 μM. Neuroglobin concentration is kept fixed at 0.1 μM. Each line shows the time-course of caspase-3 activation at the single cell level. Caspase 3 activation is normalized by the maximum. Time is measured in monte carlo (MC) simulations steps. 1 MC step = $10^{-4}$ s. Note the variation in time-scales in caspase 3 activation as the cytochrome *c* concentration is increased

### Discussion

Neuroglobin is expressed in various regions of the brain and at particularly high levels in neurons, some endocrine cell types and in retinal rod cells, reaching up to 100 μM, in association with mitochondria [5]. Neuroglobin is evolutionarily extremely conserved, with the human and frog forms being 76% identical in amino acid sequence [5].

Significant evidence has been accumulating that this protein has a physiologically highly significant role. In vivo and in vitro data have shown that neuroglobin protects cells from stroke damage, amyloid toxicity and anoxic injury [6, 8–11]. Genetic linkage analysis identified low level of neuroglobin expression in brain neurons with increased risk of Alzheimer's disease [7]. Very recent studies have further shown that neuroglobin protection of cells undergoing hyoxia-reperfusion challenge is associated with the maintenance of mitochondrial function, ATP concentration and calcium homeostasis [39, 40].

The exact mechanism by which neuroglobin achieves this protective activity is still controversial. A number of potential roles have been suggested for neuroglobin since its discovery. Initially, the search for a cellular role was influenced by the mistaken analogy to other heme proteins such as myoglobin, and thus it was suggested that neuroglobin could provide a reserve oxygen store. However, it has since been shown that oxygenated neuroglobin undergoes rapid autoxidation and has a relatively low binding affinity for oxygen [37, 41]. Therefore, neuroglobin appears to have physiological function separate from oxygen storage and transport. A role in nitric oxide (NO) scavenging was then proposed [42, 43]. Recently, neuroglobin has been shown to inhibit Rac1-mediated actin assembly and aggregation of membrane microdomains which are involved in regulation of death-signal transduction pathways [44]. Our hypothesis, that a major physiological role for neuroglobin is the interception of the mitochondrial pathway of apoptosis stems from the observation that neuroglobin reacts very rapidly with cytochrome $c$ [12, 13, 22]. Moreover, an anti-apoptotic action of neuroglobin at the immediate post-mitochondrial stage of the intrinsic pathway of apoptosis may well account for the recently reported protection of cultured neurons and primary neurons, from transgenic neuroglobin over-expressing animals, towards hypoxic challenge [39, 40].

In this work we have combined a system level, Monte Carlo signalling model of the intrinsic pathway of apoptosis with experimental studies to elucidate the functional mechanism of apoptosis inhibition by neuroglobin. We observe that neuroglobin over-expressing cells, derived from a parental neuroblastoma SH-SY5Y cell line, have significantly increased resistance to cell death induced by a BH3 mimetic HA14-1, confirming the impact of neuroglobin on the intrinsic pathway of apoptosis. Initial in vitro kinetic investigations undertaken in our laboratory identified a very rapid reaction between reduced (ferrous) neuroglobin and oxidised (ferric) cytochrome $c$ which occurs on the milli-second time scale [13]. A detailed analysis of the concentration dependence of the neuroglobin-cytochrome $c$ reaction showed that the reaction consists of two components: (1) initial protein–protein binding and (2) subsequent electron exchange between the two proteins [13]. Our systems level computational modelling reveals that both of these component reactions have the potential to prevent the activation of the intrinsic apoptotic pathway. This effect of neuroglobin is associated with the inhibition of the low probability event of apoptosome formation.

Analysis of caspase 9 activation within the reconstituted in vitro system reveals that the binding of neuroglobin to cytochrome $c$, even in the absence of any redox change, can significantly decrease activation of pro-caspase 9. Based on these experimental data we developed a second computational model to study the anti-apoptotic effects of neuroglobin, based on the assumption that neuroglobin binds to ferric cytochrome $c$ but does not reduce it. Under this assumption a higher concentration of neuroglobin is required to inhibit apoptosis. Importantly, such a concentration can still be reached in vivo, as shown in our neuroglobin over-expressing cells, and as reported earlier in neurons and some endocrine cells. Thus, while redox-reaction between neuroglobin and cytochrome $c$ may increase the anti-apoptotic activity of neuroglobin, it does not constitute an absolute prerequisite for neuroglobin action.

Mitochondria act as sentinels of diverse stress signals that emanate from other intracellular organelles, as well as a variety of environmental stress signals such as low nutrient levels, increased calcium levels, oxidative stress and intracellular neurofibrillary tangles of abnormally hyperphosphorylated tau protein that accumulates in Alzheimer's disease. In response to these signals, mitochondria become ruptured which leads to the release of pro-apoptotic molecules such as cytochrome $c$ and induction of the cascade of apoptotic events such as formation of the apoptosome and activation of the initiator caspase 9. It has been shown that neural cells can survive damage to the mitochondria as long as the formation of apoptosome is prevented. On the basis of our combined experimental and computational studies we propose that neuroglobin inhibits the intrinsic pathway of apoptosis downstream of mitochondria, where it initially binds to released cytochrome $c$ and then renders it apoptotically inactive [13]. The nature of these interactions is such that they produce a high degree of cell-to-cell stochastic fluctuations in this process. We also hypothesize that the fundamental role of neuroglobin found in neurons is to prevent accidental apoptosis occurring due to intracellular stress signals associated with normal cell functioning/physiology. We propose that the exact concentration of neuroglobin present in the cell confers a 'trigger level' to the activation of the intrinsic pathway of apoptosis. Therefore, cells, which during their normal functioning experience accidental damage to mitochondria, for example as a result of frequent calcium spikes, may well protect themselves from cytochrome $c$-induced caspase activation by expressing a high level of neuroglobin [45]. The identified action of neuroglobin in rendering cytochrome *c* apoptotically inactive provides a unique biological platform on which we can base the design of targeted therapies against a number of neural diseases.